\documentclass[aps,pra,twocolumn,amsfonts,amsmath,floatfix,showpacs,superscriptaddress]{revtex4}

\usepackage[final]{graphicx}
\usepackage{color}

\newcommand{\be}{\begin{equation}}
\newcommand{\ee}{\end{equation}}

\begin{document}

\title{Dilution and resonance enhanced repulsion in non-equilibrium
fluctuation forces}

\date{\today}

\author{Giuseppe Bimonte}
\affiliation{Dipartimento di Scienze Fisiche, Universit{\`a} di Napoli
  Federico II, Complesso Universitario MSA, Via Cintia, I-80126
  Napoli, Italy}
\affiliation{INFN Sezione di Napoli, I-80126 Napoli, Italy }

\author{Thorsten Emig}
\affiliation{Laboratoire de Physique Th\'eorique et Mod\`eles
  Statistiques, CNRS UMR 8626, B\^at.~100, Universit\'e Paris-Sud, 91405
  Orsay cedex, France}

\author{Matthias Kr\"uger}
\affiliation{Massachusetts Institute of Technology, Department of
  Physics, Cambridge, Massachusetts 02139, USA}

\author{Mehran Kardar}
\affiliation{Massachusetts Institute of Technology, Department of
 Physics, Cambridge, Massachusetts 02139, USA}

\begin{abstract}
  In equilibrium, forces induced by fluctuations of the
  electromagnetic field between electrically polarizable objects
  (microscopic {\em or} macroscopic) in vacuum are always attractive.
  The force may, however, become repulsive for microscopic particles
  coupled to thermal baths with different temperatures.  We
  demonstrate that this non-equilibrium repulsion can be realized also
  between macroscopic objects, as planar slabs, if they are kept at
  different temperatures.  It is shown that repulsion can be enhanced
  by (i) tuning of material resonances in the thermal region, and by
  (ii) reducing the dielectric contrast due to ``dilution''.  This can
  lead to stable equilibrium positions.  We discuss the realization of
  these effects for aerogels, yielding repulsion down to sub-micron
  distances at realistic porosities.
\end{abstract}


\maketitle

Forces induced by electromagnetic (EM) field fluctuations of
quantum and thermal origin act virtually between all matter that
couples to the EM field, since the interacting objects need not to
be charged \cite{parsegian_book,bordag}. Under rather general
conditions (e.~g., for non-magnetic objects in vacuum), the
Casimir potential energy does not allow for stable equilibrium
positions of the interacting objects \cite{Rahi+10}.  This can be
a practical disadvantage in systems where external
(non-fluctuation) forces cannot be applied or fine-tuned to
establish stability, especially in dynamic systems where the
distance and hence the Casimir force changes in time.
Nano-mechanical devices with closely spaced components fall into
this class of systems \cite{Serry+95}. Repulsive Casimir forces
are known to exist if the space between the objects is filled by a
dielectric with suitable contrast \cite{DLP61}, but this is
impractical in many situations.

Repulsion can occur also in response to the preparation of
particles in distinct internal states, e.g. by optical excitation
or coupling to heat baths of different temperatures. 
In particular, Cohen and Mukamel \cite{cohen} predicted a
non-equilibrium repulsive force between molecules generated
by suitable detuning of the resonance frequencies. 
This suggests that for macroscopic objects held at different temperatures
similar repulsive effects should exist close to material
resonances in the thermal region. However, for macroscopic
condensates the dielectric contrast is usually strong and
non-equilibrium effects are comparatively less significant than the
equilibrium attraction. One should thus focus on
sufficiently optically diluted materials to generate resonant
Casimir repulsive forces between macroscopic bodies.
The general formalism for dealing with non-equilibrium fluctuation effects 
between macroscopic bodies has been developed recently in
\cite{pita,bimonte,Krueger+11,antezza,Krueger+111}.

The aim of the present work is to explore theoretically if and to what
extent the above expectation can be realized and whether it can lead to stable
equilibrium positions. 
We consider two dielectric slabs held at different temperatures, and
computed the pressure between them using the non-equilibrium 
extension of the Casimir--Lifshitz formula \cite{pita}.
Employing a Lorentz-Drude dielectric response, we find that resonances
and optical dilution indeed amplify repulsion sufficiently so that the
total interaction can become repulsive, and equilibrium positions exist. 
We show that aerogels could be used to realize resonant
repulsion in practice since porosity can be used to tune reflectivity
and resonances. The use of aerogels was indeed
previously proposed to reduce the Casimir force \cite{esquivel07}, but
not to generate repulsion.

We consider two infinite parallel planar slabs ${\cal S}^{\alpha},\;
\alpha=1,2$, each consisting of a non-magnetic dielectric layer of
thickness $\delta$ and permittivity $\epsilon_{\alpha}(\omega)$,
deposited on a thick glass substrate of dielectric permittivity
$\epsilon_{\rm sub}(\omega)$.  
The slabs are held at temperatures $T_1$ and $T_2$, 
separated by a (vacuum) gap of width $a$.
The Casimir pressure acting on the inside faces of the
  plates is given by \cite{pita}
\begin{align} 
P_{\rm neq}(T_1,T_2,a) &={\bar P}_{{\rm eq}}(T_1,T_2,a)+\notag
 \Delta
P_{{\rm neq}}(T_1,T_2,a)\\&+\frac{2\sigma}{3c}(T_1^4+T_2^4)\;,
\label{prestotn}
\end{align}
where $\sigma$ is the Stefan-Boltzmann constant.  
The last term in this equation is simply the classical thermal radiation
result, which is independent of distance and material properties.
In Eq.~\eqref{prestotn}, ${\bar  P}_{{\rm eq}}(T_1,T_2,a)=[P_{{\rm eq}}(T_1,a)+P_{{\rm
    eq}}(T_2,a)]/2$ denotes the average of the equilibrium Casimir
pressures at $T_1$ and $T_2$; with $P_{{\rm eq}}(T,a)$  given by the
Lifshitz formula \cite{lifs}, which for brevity is not reproduced here.
The novel non-equilibrium contributions are captured by the
term $\Delta P_{{\rm neq}}(T_1,T_2,a)$, 
which vanishes for $T_1=T_2$ (it also contains a distance
independent part). Upon decomposing $\Delta P_{{\rm
    neq}}=\Delta P_{{\rm neq}}^{\rm PW}+\Delta P_{{\rm neq}}^{\rm EW}$
into propagating waves (PW) and evanescent waves (EW), 
one finds~\footnote{We use a sign convention opposite to Ref.~\cite{pita},
such that negative pressures represent attraction between the slabs.}
\begin{align}
& \Delta P_{{\rm neq}}^{\rm PW} =\frac{\hbar}{4
\pi^2}\sum_{P=M,N}\int_0^{\infty} \!\!\!d \omega \,\left[n(T_1)-
n(T_2) \right]\nonumber\\
& \times \int_0^{\omega/c} d k_{\perp}
k_{\perp}\,k_z
\frac{|r_{P}^{(2)}|^2-|r_{P}^{(1)}|^2}{|D_{P}|^2} \;,\label{noneqPW} \\
& \Delta P_{{\rm neq}}^{\rm EW}  =-\frac{\hbar}{2
\pi^2}\sum_{P=M,N}\int_0^{\infty} \!\!\!d \omega \, \left[n(T_1)-
n(T_2) \right]\int_{\omega/c}^{\infty}\!\!\!\!d
k_{\perp} k_{\perp}\nonumber\\
& \times\,{\rm Im}(k_z) e^{-2 a {\rm Im}(k_z) } \frac{{\rm
Im}(r_{P}^{(1)}){\rm Re}(r_{P}^{(2)})-{\rm Re}(r_{P}^{(1)}){\rm
Im}(r_{P}^{(2)})}{|D_{P}|^2} \;,\label{neqEW}
\end{align}
where  $n(T)=[\exp(\hbar \omega/k_B
T)-1]^{-1}$, $k_z={\sqrt{\omega^2/c^2-k_{\perp}^2}}$, 
$D_{P}=1-r_{P}^{(1)}r_{P}^{(2)}\,\exp(2 i k_z\,a)$,
and $P=M,N$ for the two polarizations. The reflection
coefficients $r^{(\alpha)}_{P}(\omega,k_{\perp})$ are given by the
well-known formulae for a two-layer slab \be
r^{(\alpha)}_{P}=\frac{r_{P}(1,\epsilon_{\alpha})+r_{P}(\epsilon_{\alpha},\epsilon_{\rm
sub})\,e^{2i \delta
q_{\alpha}}}{1+r_{P}(1,\epsilon_{\alpha})\,r_{P}(\epsilon_{\alpha},\epsilon_{\rm
sub})\,e^{2i \delta q_{\alpha}}} \;,\ee where
$r_{P}(\epsilon_a,\epsilon_b)$ are the Fresnel reflection
coefficients
\be
r_{N}(\epsilon_a,\epsilon_b)=\frac{\epsilon_{b}(\omega)q_a(\omega,k_{\perp})-\epsilon_{a}(\omega)q_b(\omega,k_{\perp})
}{\epsilon_{a}(\omega)q_b(\omega,k_{\perp})+\epsilon_{b}(\omega)q_a(\omega,k_{\perp}
}\;, \label{ppol} \ee
where $q_a=\sqrt{\epsilon_a(\omega) \omega^2/c^2-k_{\perp}^2}$,
and $r_{M}(\epsilon_a,\epsilon_b)$ is obtained by replacing the
$\epsilon_a$ and $\epsilon_b$ in 
Eq.~(\ref{ppol})  by one (but not in $q_a$).
There is also an external pressure acting on the outside face of each plate
which depends on the reflectivity of this face
and the temperature of the environment $T_{\rm env}$. In our
numerical computations we assume that the external face of the
glass substrate has been blackened, in which case the total
pressure on plate $\alpha$ is given by
\begin{equation}
{\tilde P}^{(\alpha)} (T_1,T_2,T_{\rm env},a)=P_{\rm neq}(T_1,T_2,a)-\frac{2\sigma}{3c}(T_\alpha^4+T_{\rm env}^4).
\end{equation}
It is worth emphasizing that for $T_1\not= T_2$,
${\tilde P}^{(1)} (T_1,T_2,T_{\rm
  env},a) \neq {\tilde P}^{(2)}(T_1,T_2,T_{\rm env},a)$. 
Also, in equilibrium all distance independent terms vanish.

The structure of the non-equilibrium force in Eq.~(\ref{prestotn})
suggests that repulsion should exist out of thermal equilibrium. While
the first term  ${\bar P}_{{\rm  eq}}(T_1,T_2,a)$  in Eq.~(\ref{prestotn})
is bound to be attractive, this is not so for the second term $\Delta P_{{\rm neq}}(T_1,T_2,a)$. 
As it can be
seen from Eqs.~(\ref{noneqPW}) and (\ref{neqEW}), the quantity
$\Delta P_{{\rm neq}}(T_1,T_2,a)$ changes sign if the temperatures
$T_1$ and $T_2$ are exchanged, and therefore its sign can be
reversed by simply switching the temperatures of the plates. One
notes also that $\Delta P_{{\rm neq}}(T_1,T_2,a)$ is {\it
antisymmetric} under the exchange $r_P^{(1)} \leftrightarrow
r_P^{(2)}$, and vanishes for $r_P^{(1)} = r_P^{(2)}$. Therefore in
order to take advantage of this term to control the sign of the
Casimir force it is mandatory to consider plates made of {\it
different} materials. For real materials, both ${\bar
    P}_{{\rm eq}}(T_1,T_2,a)$ and $\Delta{P}_{{\rm neq}}(T_1,T_2,a)$
  diverge as $a^{-3}$ if $a\to0$. In the following, we show that the
  sign of this asymptotic behavior can be made repulsive in certain
  cases.

Motivated by the resonance-induced repulsion for microscopic particles, we
consider electric permittivities $\epsilon_{\alpha}(\omega)$ of
the dielectric layers described by a Lorentz-Drude type model, as
 \be
\epsilon_{\alpha}(\omega)=1+\frac{C_{\alpha}\,\omega_{\alpha}^2}{\omega_{\alpha}^2-\omega^2-i
\gamma_{\alpha}
\omega}+\frac{D_{\alpha}\,\Omega_{\alpha}^2}{\Omega_{\alpha}^2-\omega^2-i
\Gamma_{\alpha} \omega}\;.\label{model} \ee
(An analogous two-oscillator model was used also for the
permittivity $\epsilon_{\rm sub}$ of the glass substrate, with the
parameters quoted in \cite{bordag}.) The first oscillator term
($\sim C_{\alpha}$) describes low-lying excitations of the materials;
such low-lying polariton excitations in numerous dielectrics 
account for sharp peaks in their dielectric functions in the far infrared region. 
Typical values for the
resonance and relaxation frequencies are
$\omega_{\alpha}=10^{13}-10^{14}$ rad/sec and
$\gamma_{\alpha}=10^{11}- 10^{12}$ rad/sec~\cite{Kittel}.
The second oscillator term in  Eq.~(\ref{model}), proportional to
$D_{\alpha}$, describes the contribution of core electrons.
Excitation energies of core electrons  are much larger, and
characteristic values of $\Omega_{\alpha}$ are in the range
$10^{15}- 10^{16}$ rad/sec.  At and around
room temperature core electrons are not thermally excited
(for $T=300$K the characteristic thermal frequency $\omega_T=k_B
T/\hbar$ is 3.9 $\times 10^{13}$ rad/sec) and therefore their
contribution to the thermal Casimir force $\Delta P_{\rm neq}$ is
very small. However core electrons  are important, as   they strongly
contribute  to the average equilibrium Casimir force ${\bar
P}_{\rm eq}$, especially at sub-micron separations.

Since the thermal Casimir force between two macroscopic slabs
should reduce in the dilute limit to the pairwise interaction
between their atoms, one expects that the resonant phenomena
reported in Ref.~\cite{cohen} should be recovered if the material of
the plates is sufficiently diluted. In order to determine how
large a dilution is necessary for this to happen, we investigated
the behavior of the Casimir force under a rescaling of  the
amplitudes of the resonance peaks,
  i.e. $C_{\alpha} \to C_{\alpha}/\tau$,  $D_{\alpha} \to D_{\alpha}/\tau$
  by a  overall optical dilution
parameter $\tau \geq 1$.

\begin{figure}
\includegraphics{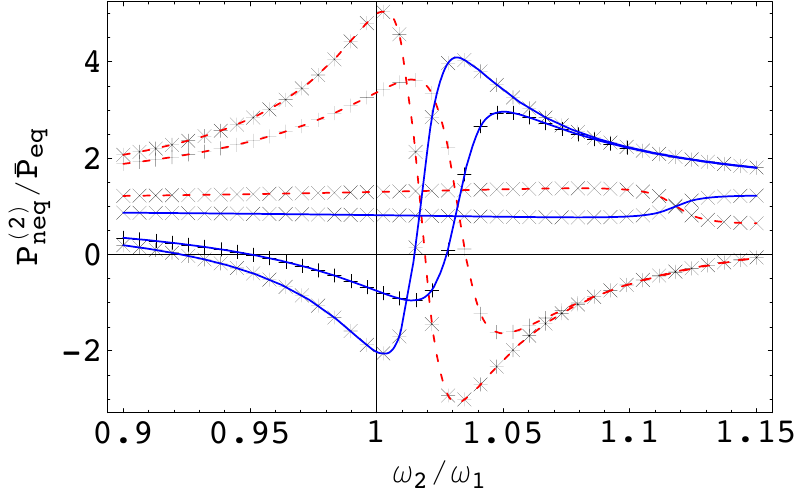}
\caption{\label{Fig1}   The normalized non-equilibrium Casimir
pressure $P_{\rm neq}^{(2)}/{\bar P}_{\rm eq}$ on  slab 2 as a function of
the ratio of the resonance frequencies $\omega_2/\omega_1$, for
$a=300$ nm, $\delta=5\,\mu$m, $C_1=3$, $C_2=1.5$, $D_1=1$, and
$D_2=0.5$. The red dashed curves are for $T_1=T_{\rm env}=300$K and
$T_2=600$K, while blue solid curves are for $T_1=T_{\rm env}=300$K and $T_2=150$K. 
Three values of the dilution parameter $\tau$
are displayed: $\tau=1$ ($\times$), $\tau=10$ ($+$) and $\tau=20$
($\ast$). Negative values of $ P_{\rm neq}^{(2)}/{\bar P}_{\rm
eq}$ correspond to repulsion.}
\end{figure}

We next report on numerical results based upon the above model for dielectrics,
in which we set $\omega_1=10^{13}$, $\gamma_1=
\gamma_2=10^{11}$, $\Omega_1=\Omega_2=10^{16}$, and
$\Gamma_1=\Gamma_2=5 \times 10^{14}$ (all in rad/sec).  In
Fig.~\ref{Fig1} we plot the non-equilibrium normalized Casimir
pressure $ P_{\rm neq}^{(2)}/{\bar P}_{\rm eq}$ on slab 2
versus the ratio of the resonance frequencies $\omega_2/\omega_1$, 
for $a=300$ nm, $\delta=5\,\mu$m,
$C_1=3$, $C_2=1.5$, $D_1=1$, and $D_2=0.5$. 
(Negative values of $ P_{\rm neq}^{(2)}/{\bar P}_{\rm eq}$ correspond to repulsion.) 
For these parameters the non-equilibrium force is dominated by 
evanescent waves, whose skin depth is comparable to the separation $a$. 
This implies that for separations $a \ll \delta$ the force is practically independent of
$\delta$, but for $a \gtrsim \delta$ the features of the substrate
influence the magnitude of the force significantly.
The red dashed curves in Fig.~\ref{Fig1} are for $T_1=T_{\rm env}=300$K and
$T_2=600$K, while blue solid curves are for $T_1=T_{\rm env}=300$K and $T_2=150$K. 
Three values of the dilution parameter $\tau$
are displayed: $\tau=1$ ($\times$), $\tau=10$ (+) and $\tau=20$ ($\ast$). 
We see that for sufficiently high dilution, the Casimir
force is strongly dependent on the ratio $\omega_2/\omega_1$ of
the resonance frequencies of the plates, displaying features analogous
to those reported in Ref.~\cite{cohen} for the van der Waals
interaction between two `atoms' coupled to baths at different temperatures. 
In particular, the Casimir force becomes repulsive if this ratio  is suitably
tuned to a value close to unity. 
We also find that the equilibrium force is
rather insensitive to the tuning of resonances, and that the total force
hardly depends on $\Omega_1/\Omega_2$.

\begin{figure}
\includegraphics{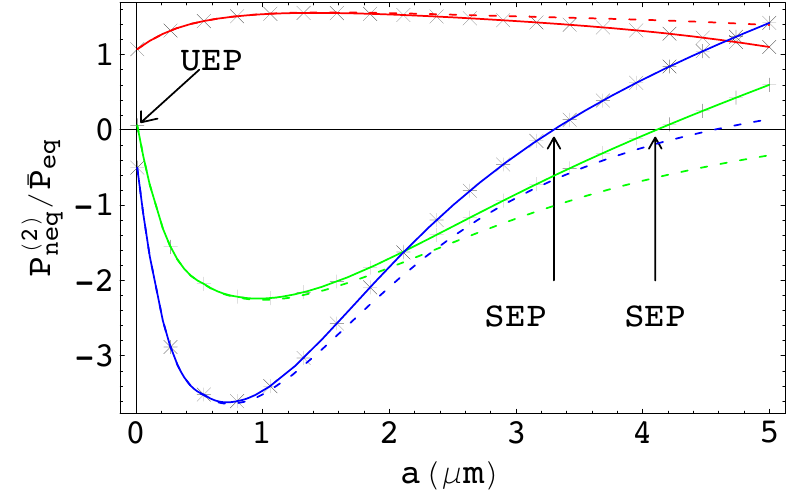}
\caption{\label{Fig2} The non-equilibrium normalized Casimir pressure
$P_{\rm neq}^{(2)}/{\bar P}_{\rm eq}$ on slab 2 as a function of plate
separation (in microns) for $\delta=5\,\mu$m, $T_1=T_{\rm
env}=300$K, $T_2=600$K, $\tau=1$ and $\omega_2/\omega_1=1.1$
($\times$), $\tau=10$ and $\omega_2/\omega_1=1.05$ (+), for
$\tau=20$ and $\omega_2/\omega_1=1.03$ ($\ast$). All other
parameters are same as in Fig.~\ref{Fig1}. Dashed curves do not
include the distance independent part of the pressure. Stable
equilibrium positions are marked as SEP, unstable ones as UEP. 
The curve for $\tau=20$ approaches a negative values $a \rightarrow 0$.}
\end{figure}

In Fig.~\ref{Fig2} we plot the dependence of the non-equilibrium normalized Casimir
pressure $ P_{\rm neq}^{(2)}/{\bar P}_{\rm eq}$ on slab 2, as a function of
plate separation $a$, at $T_1=T_{\rm env}=300$K, $T_2=600$K, for
$\tau=1$ and $\omega_2/\omega_1=1.1$ ($\times$); 
$\tau=10$ and $\omega_2/\omega_1=1.05$ (+); 
$\tau=20$ and $\omega_2/\omega_1=1.04$ ($\ast$), 
all other parameters being same as in Fig.~\ref{Fig1}. 
The dashed curves do not include the distance independent part of the pressure,
and are included to indicate that this component also changes sign upon dilution. 
We see that without dilution ($\tau=1$) the non-equilibrium Casimir
force is attractive for all displayed separations. 
By contrast, for large enough dilutions and for suitable values of $\omega_2/\omega_1$, 
the force becomes repulsive in a wide range of separations.
The curve for $\tau=10$ exhibits two points of zero force, one at $a=15$~nm and
another for $a=4.1$~$\mu$m, corresponding to an unstable equilibrium
point (UEP) and a stable equilibrium point (SEP), respectively. 
For $\tau=20$ there is a single turning point corresponding to
a SEP at $a=3.3$~$\mu$m. There is no UEP point in this case as
the ratio $ P_{\rm neq}^{(2)}/{\bar P}_{\rm eq}$ approaches a negative value 
in the limit $a \rightarrow 0$, signifying that repulsion persists for arbitrarily 
small plate separations less than the SEP.

As a means of achieving the dilution levels required  to observe
the resonance phenomena described above, we consider aerogels: highly porous
materials fabricated by sol-gel techniques, starting from a variety of materials 
such as ${\rm SiO}_2$, carbon, ${\rm Al}_2{\rm O}_3$, platinum etc. 
Aerogels with levels of porosity exceeding 99$\%$ can be realized nowadays~\cite{Pierre}.   
In order to study the Casimir force between two aerogel plates we need an expression
for the dielectric function ${\hat \epsilon}(\omega)$ of the aerogel, 
valid in the wide range of frequencies relevant for the Casimir effect. 
For separations $a$ larger than the pore  size (typically of the order of 100~nm or less),
an effective medium approach can be used in which the aerogel 
permittivity $\hat{\epsilon} (\omega)$ is obtained from the 
Maxwell-Garnett equation~\cite{Choy_book} as
 \be
\frac{\hat{\epsilon}-1}{\hat{\epsilon}+2}=(1-\phi)\frac{{\epsilon}-1}{{\epsilon}+2}\;.
\label{CM}
\ee
Here, $\epsilon(\omega)$ is the permittivity of the solid
 fraction of the aerogel, and  $0\le \phi \le 1$ is the porosity.
Equation~\eqref{CM} is justified if the solid fraction is well
separated by the host material (air), i.e., when $\phi$ is
sufficiently close to one. Using again  a model of the form in Eq.~(\ref{model}) for $\epsilon_{\alpha}(\omega)$, one finds that $\hat{\epsilon}_{\alpha}(\omega)$
has a   resonance at the frequency 
 \be
{\hat \omega}_{\alpha} \approx
\left(1+\frac{\phi_{\alpha}\,C_{\alpha}}{\phi_{\alpha}
\,D_{\alpha}+3}\right)^{1/2}\, {\omega}_{\alpha}\;, \label{resaer}
\ee where we assumed $\omega_{\alpha} \ll \Omega_{\alpha}$.
According to Eq.~(\ref{resaer}), the frequency of the aerogel
resonance is blue-shifted with respect to
${\omega}_{\alpha}$, and as $\phi_{\alpha}$ is varied from zero to
one, ${\hat \omega}_{\alpha}$ sweeps  the range from
$\omega_{\alpha}$ to $[1+C_{\alpha}/(D_{\alpha}+3)]^{1/2}
\omega_{\alpha}$. The dependence of the resonance frequency
 $\hat{\omega}_{\alpha}$ on the porosity is welcome,
because it gives us the possibility of tuning the ratio of the
resonance frequencies for the two plates by simply choosing
appropriate values for the porosities $\phi_1$ and $\phi_2$ of the plates.

\begin{figure}
\includegraphics{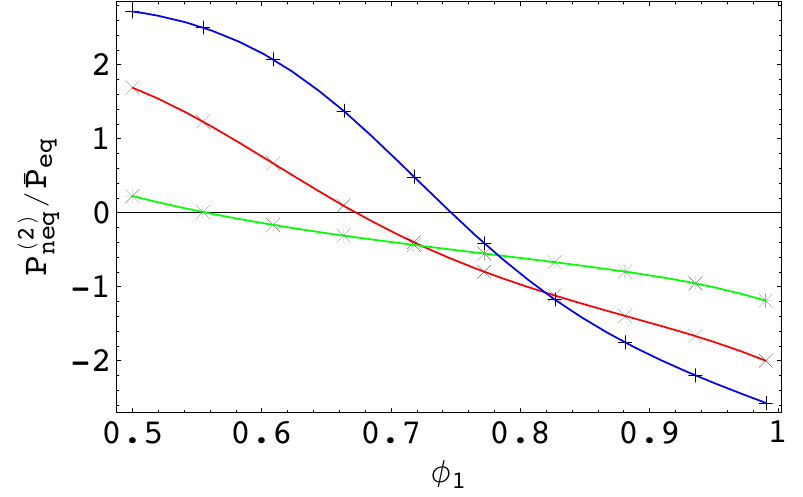}
\caption{\label{Fig3}   The pressure ratio $P_{\rm neq}^{(2)}/{\bar P}_{\rm eq}$ 
for two aerogel  layers ($\delta=5\, \mu$m) on a glass substrate
as a function of the porosity $\phi_1$ of plate one. The
porosity $\phi_2$ of plate two is 0.95 (+), 0.9 ($\times$) and 0.8
($\ast$). The dielectric functions $\epsilon_{\alpha}(\omega)$ of
the host materials are as in Eq.~(\ref{model}), with $C_1=1$,
$C_2=3$, $D_1=D_2=0.5$, $\omega_2/\omega_1$=0.84.  All curves are
for a separation of 200~nm and for $T_1=T_{\rm env}=300$~K,
$T_2=600$~K.  }
\end{figure}

As an example, we consider  two aerogel layers of thickness
$\delta=5\mu$m, deposited on a thick glass substrate, with a
blackened outer surface. The dielectric functions
$\epsilon_{\alpha}(\omega)$ of the host materials are as in Eq.~(\ref{model}), 
with $C_1=1$, $C_2=3$, $D_1=D_2=0.5$, $\omega_2/\omega_1$=0.84. 
In Fig.~\ref{Fig3} we plot the ratio
$P_{\rm neq}^{(2)}/{\bar P}_{\rm eq}$  as a function of the porosity
$\phi_1$ of the first plate. The porosity $\phi_2$ of the second
plate   is   0.95 (+), 0.9 ($\times$) and 0.8 ($\ast$).  All
curves in Fig.~\ref{Fig3} are for a separation of 200~nm, at
$T_1=T_{\rm env}=300$~K and $T_2=600$~K. 
The force can indeed be made repulsive by suitably adjusting the porosities of the two
plates. In absolute terms, the Casimir force is small; e.g.
for $\phi_1=0.77$ and $\phi_2=0.8$ we find ${P}_{\rm neq}^{(2)}=0.55 \times 10^{-3}$ Pa.

\begin{figure}
\includegraphics{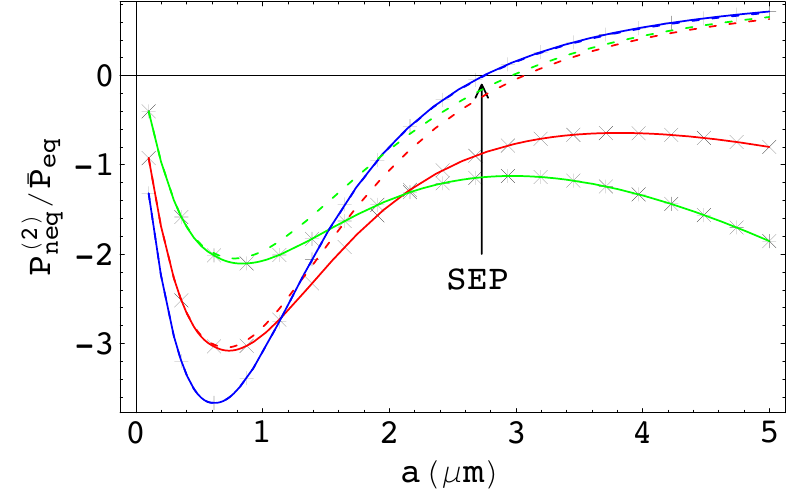}
\caption{\label{Fig4} The pressure ratio $P_{\rm neq}^{(2)}/{\bar P}_{\rm eq}$
for two aerogel  layers ($\delta=5\, \mu$m) on a glass substrate
as a function of plate separation (in microns), for  $\phi_1=0.95$ and
$\phi_2$=0.95 (+),  0.9 ($\times$)  and 0.8 ($\ast$). All
parameters for the aerogel plates are same as in Fig.~\ref{Fig3}.
The dashed lines do not include the separation independent part of
the pressure. The stable equilibrium position is marked by SEP.}
\end{figure}

In Fig. ~\ref{Fig4} we plot dependence of the ratio $P_{\rm neq}^{(2)}/{\bar P}_{\rm eq}$ 
on the separation (in microns), for $\phi_1=0.95$ and
$\phi_2$=0.95 (+), $\phi_2$=0.9 ($\times$) and $\phi_2$=0.8
($\ast$). All parameters for the aerogel plates are the same as in
Fig.~\ref{Fig3}. The dashed lines do not include the separation
independent part of the pressure. We find that for
$\phi_1=\phi_2=0.95$ a stable equilibrium point exists (marked as SEP
in Fig.~\ref{Fig4}).

While fluctuation induced forces are generally attractive, repulsive forces
can be obtained between atoms prepared in different excited states.
Coupling of atoms to thermal baths at different temperatures can in principle
produce population of states that lead to repulsion.
We show that a similar non-equilibrium repulsive contribution to pressure
also arises from the interplay of resonances for two macroscopic slabs
held at appropriate distinct temperatures.
However, to observe a net repulsion between slabs one must overcome
an ever present attractive force  arising from  the dielectric contrast of the 
condensed bodies from the intervening vacuum. 
The latter can be reduced by dilution, and we have shown that aerogels
provide a material where a net repulsion at sub-micron separations
can be achieved, leading to a stable equilibrium point.

This research was supported by the ESF Research Network CASIMIR,
DARPA contract No.~S-000354, DFG grant No.~KR 3844/1-1 and NSF
Grant No.~DMR-08-03315.
We have benefitted from discussions with R.L. Jaffe, M.F. Maghrebi,
U. Mohideen, and R. Zandi.

\end{document}